\documentstyle[12pt,epsf]{article}
\begin{document}
\hfill DESY 97-193
\begin{center}
{\sc Nonabelian Nature of Asymmetry for the $B_c$ Meson Production in
Gluon-Photon Interaction}\\

\vspace*{5mm}
A.V.Berezhnoy, V.V.Kiselev, A.K.Likhoded\\
\vspace*{0mm}~\\
Institute for High Energy Physics,\\
Protvino, Moscow Region, 142284, Russia\\
E-mail: kiselev@mx.ihep.su~~~~  Fax: +7-(095)-2302337
\end{center}

\begin{abstract}
Calculations of cross-sections for the gluon-photon production of
$B_c$ and $B_c^*$  mesons as well as for the production in 
$ep$-interactions at HERA are performed
in the leading order of perturbative QCD. We show that the nonabelian
structure of QCD leads to an essential forward-backward asymmetry in these
processes.
\end{abstract}

\section{Introduction}
A significant progress in the understanding of the
mechanisms for the
production of heavy quarkonium consisting of two heavy quarks with
different flavors, for instance,
$(\bar bc)$-system, has been achieved during past
three years \cite{ufn,prspec,zp1,zp2}.
One clarified that for the $B_c$ production in $e^+e^-$-annihilation,
the differential cross-section has the simple factorized form in the
limit of high energies ($M^2_{B_c}/s \ll 1$) \cite{frag},
\begin{equation}
\frac{d^2\sigma_{B_c}}{dzdp_T}=
\left. \frac{d\sigma_{b \bar b}}{dk_T}
\right|_{k_T=\frac{p_T}{z}}\cdot \frac{D_{\bar b \to B_c}(z)}{z},
\end{equation}
where $z=2E_{B_c}/\sqrt{s}$,
$\sigma_{b \bar b}$ is the Born cross-section for the $b \bar b$-pair
production, $k_T$ is the transverse momentum of $\bar b$-quark,
$p_T$ is the transverse momentum of $B_c$ meson, and $D_{\bar b \to B_c}(z)$
is interpreted as the fragmentation function of $\bar b \to B_c+X$.
In the language of Feynmann diagrams, the latter fact means that in the
high energy limit one can conveniently choose a gauge, namely
the axial gauge, with the unit $n$ vector directed along the
four-momentum of the $b$-quark, so that  the contribution of 
the diagrams where
the $c \bar c$-pair is produced off the $\bar b$-quark line
is dominant.
Using this fact, one analytically calculated the fragmentation functions
for the bound states in $S$-, $P$- and $D$-waves \cite{frag}.

The production of $(\bar b c)$-quarkonium in hadronic, photonic and 
photon-gluon interactions occurs in a more complex way and it can not
be described by eq. (1) in the entire kinematical region even at
high energies  $M^2_{B_c}/s \ll 1$. The number of diagrams for the 
mentioned processes is greater than that in $e^+e^-$-annihilation
(one has 36 diagrams for the gluonic interactions, 20 for the photonic
collisions and 24 for the photon-gluon subprocess).
So, one can find no gauge to restrict the number of dominant diagrams by 
those  of the fragmentation kinds , i.e., by diagrams like 15 or 18 in Fig. 1,
where the tree process $g \gamma \to B_c +X$ is presented. The origin
of the problem is that the recombination diagrams also contribute to the 
cross-section, for example, diagram  5 in Fig. 1 is of this
kind. This contribution is essential up to the transverse momentum
values equal to 35 GeV. At $p_T>35$ GeV, the regime of fragmentation (1)
becomes dominant, as one must expect by the theorem on the hard process
factorization \cite{soper}. However, the latter fact underlines
the calculations of the contributions, determined by the complete set of
diagrams.
The complete perturbative calculations can lead to a possible interference
of various contributions in some kinematical regions. So, detailed analysis
for the gluonic production of $B_c$ mesons at high energies was performed by us
\cite{pert}. We found that the effect of destructive interference
between the fragmentation diagrams and the 
ones containing the three-gluon vertex is important,
so that in the region of $z$ close to $1$ and at moderate $p_T$ values
 the complete set of diagrams leads to smaller 
cross-section in comparison with the 
one calculated according to (1) \cite{pert}. 
We also found that for the
photon-photon interaction the interference is absent.
These investigations lead us to the conclusion that 
destructive interference
takes place if the interaction possesses the nonabelian character. Therefore,
it is interesting to study the production of $B_c$ meson 
in the gluon-photon interactions,
wherein the mentioned interference would result in the forward-backward
asymmetry, which is described in this paper.

Unfortunately, a straightforward comparison of experimental
measurements with the 
fragmentation picture is problematic since the condition
$M_{B_c}^2/s\ll 1$ is generally not satisfied in partonic subprocesses, 
and the dominant
contribution to the cross-section is integrated over the region
close to the threshold of the meson production in the subprocess.
In this work the cross-section of $B_c$ production in the HERA experiments
will be presented. Despite the dominance of the kinematical region 
near the subprocess threshold, the effect
caused by the presence of the three-gluon vertices, i.e. the nonabelian texture
of QCD, turns out to be significant.

\section{Calculation technique}
The amplitude for the $B_c$ meson production, $A^{SJj_z}$, can be 
expressed through 
the amplitude of four free quarks production, $T^{Ss_z}(p_i,k({\bf q}))$,
and the orbital wave function of the $B_c$ meson, $\Psi^{Ll_z}({\bf q})$, 
in the meson rest frame as
\begin{equation}
A^{SJj_z}=\int T^{Ss_z}(p_i,k({\bf q}))\cdot 
\left (\Psi^{Ll_z}({\bf q}) \right )^* \cdot
C^{Jj_z}_{s_zl_z} \frac{d^3 {\bf q}}{{(2\pi)}^3},
\label{int}
\end{equation}
where $J$ and $j_z$ are the total spin of the meson and its projection on $z$ 
axis in the $B_c$ rest frame, respectively; $L$ and $l_z$ are the orbital
momentum and its projection; $S$ and $s_z$ are the sum of quark spins and its
projection; $C^{Jj_z}_{s_zl_z}$ are the Clebsch-Gordan coefficients;
$p_i$ are four-momenta of $B_c$, $b$ and $\bar c$,
${\bf q}$ is the three-momentum of $\bar b$ quark in the $B_c$ meson rest
frame; $k({\bf q})$ is the four-momentum, obtained from the
four-momentum $(0,{\bf q})$ by the Lorentz transformation from the
$B_c$ rest frame to the system, where the calculation of
$T^{Ss_z}(p_i,k({\bf q}))$ is performed. Then, the four-momenta
of $\bar b$ and $c$ quarks, forming the $B_c$ meson, will be determined 
by the following formulae with the accuracy up to $|{\bf q}|^2$ terms
\begin{equation}
\begin{array}{c}
p_{\bar b}=\frac{m_b}{M}P_{B_c}+k({\bf q}), \\
p_{c}=\frac{m_c}{M}P_{B_c}-k({\bf q}),
\end{array}
\label{mom}
\end{equation}
where $m_b$ and $m_c$ are the quark masses, $M=m_b+m_c$,
and $P_{B_c}$ is the $B_c$ momentum. Let us  note that for the 
$S$-wave states it is
enough to take into account only terms independent of ${\bf q}$.

The 24 diagrams, contributing to the gluon-photon
production of $B_c$ are shown in Fig. 1.
The product of spinors $v_{\bar b} \bar u_c$, corresponding to the
$\bar b$ and $c$ quarks in the amplitude $T^{Ss_z}(p_i,k({\bf q}))$ of
eq.(\ref{int}), should be substituted by the projection operator
\begin{equation}
{\cal P} (\Gamma )=\sqrt{M} \left (\frac{\frac{m_b}{M} \hat P_{B_c}
+\hat k-m_b}{2m_b}
\right ) \Gamma \left (\frac{\frac{m_c}{M}\hat P_{B_c}-\hat k+m_c}{2m_c}
\right ),
\end{equation}
where $\Gamma=\gamma^5$ for $S=0$, or $\Gamma=\hat \varepsilon^*(P_{B_c},s_z)$
for $S=1$, where $\varepsilon(P_{B_c},s_z)$ is the polarization vector for the
spin-triplet state.

The fact that the $\bar b c$-system is produced in the colour-singlet state
simplifies the texture of colour matrix. The latter has the only
nonzero eigenvalue for the given process, so that one can significantly
reduce the time for the computation of the matrix element.

The $S$-wave production amplitude can be written down as
\begin{equation}
A^{Ss_z}=iR_S(0)\sqrt{\frac{2M}{2m_b2m_c}}
\sqrt{\frac{1}{4\pi}}
\left (T^{Ss_z}\left (p_i,k({\bf q}=0)\right )\right ),
\label{main0}
\end{equation}
where $R_S(0)$ is the radial wave function at the origin, so that
$\Psi (0)= R(0)/\sqrt{4\pi}$ and
$$
R_S(0) = \sqrt{\frac{\pi M}{3}}\; \tilde f_{B_c},
$$
and the quantity $\tilde f_{B_c}$ is related to the leptonic decay constants
of pseudoscalar and vector $B_c$ states
\begin{eqnarray}
\langle 0| J_\mu(0)|V\rangle & = & i f_V M_V\; \epsilon_\mu\;, \nonumber\\
\langle 0| J_{5\mu}(0)|P\rangle & = & i f_P p_\mu\;, \nonumber
\end{eqnarray}
where $J_{\mu}(x)$ and $J_{5\mu}(x)$ are the vector and axial-vector currents
of the constituent quarks. In the lowest order in $\alpha_s$ 
one has  $\tilde f=f_{\rm V}, f_{\rm P}$.
The account for hard gluon corrections
in the first order in $\alpha_s$ \cite{bra,int} results in
\begin{eqnarray}
\tilde f & = & f_V\; \bigg[1 - \frac{\alpha_s^H}{\pi}
\biggl(\frac{m_2-m_1}{m_2+m_1}\ln\frac{m_2}{m_1} -\frac{8}{3}\biggr)\bigg]\;,\\
\tilde f & = & f_P\; \bigg[1 - \frac{\alpha_s^H}{\pi}
\biggl(\frac{m_2-m_1}{m_2+m_1}\ln\frac{m_2}{m_1} - 2\biggr)\bigg]\;,
\end{eqnarray}
where $m_{1,2}$ are the masses of the quarks composing the quarkonium. For the
vector currents of quarks with equal masses, the BLM procedure of scale 
fixing in the "running" coupling constant of QCD \cite{blm} gives 
(see ref. \cite{vol})
$$
\alpha_s^H = \alpha_s^{\overline{\rm MS}}(e^{-11/12}m_Q^2)\;.
$$
The estimates of the $\tilde f_{B_c}$ value within the potential models
and QCD sum rules are somewhat uncertain, 
$\tilde f_{B_c}=460 \pm 100$ MeV \cite{prspec}.
Recent estimates within the QCD sum rules and 
in the framework of lattice computations give
$f_{B_c}= 385\pm 25$ MeV and $f_{B_c}=395(2)$ MeV, respectively.
Taking into account radiative corrections, the value of
$\tilde f_{B_c}$ is about $470$ MeV \cite{hs}, which will be used in the
following estimates. Since the cross-section is proportional to
$\tilde f_{B_c}^2$, the corresponding value of the cross-section 
for a different value of
$\tilde f_{B_c}$ can be obtained by  rescaling with the
multiplicative factor $(\tilde f_{B_c}/\tilde f_{B_c}')^2$.

\section{Results of calculations}
In the leading order of perturbation theory, the total cross-section for 
the photon-gluon production of $B_c$ is proportional to
$
\sigma \sim \frac{\alpha \alpha_s^3}{m_b^2m_c^3}\;|\Psi(0)|^2
$, where $|\Psi(0)|$, $\alpha_s$ and masses of quarks are
free parameters of the model. The calculations are performed with the following
values of the parameters
\begin{equation}
\begin{array}{lll}
\alpha &=& 1/133,\\
\alpha_s &=& 0.24,\\
m_b &=& 4.8\  {\rm GeV},\\
m_c &=& 1.5\ {\rm GeV},\\
\tilde f_{B_c} &=& 470\ {\rm MeV}.
\end{array}
\end{equation}
The results for the gluon-photon production of $S$-wave
levels in the $(\bar b c)$-quarkonium ($B_c$ and $B_c^*$) are 
presented in Fig. 2 and Tab. 1. For the sake of comparison, 
the predictions of the fragmentation mechanism are also shown in Fig. 2.
One can see that the fragmentation mechanism underestimates
the cross-section, much the same way as in the photonic and 
gluonic interactions producing $B_c$ (or $B_c^*$) \cite{zp1,zp2}.

The distributions $d\sigma /dz$ for both the $B_c$ meson moving in the
gluon hemisphere (i.e. at $p_z>0$, where $p_z$ is the projection of the
momentum on the beam axis in the c.m.s. of the colliding particles) and
$B_c$ moving in the photon hemisphere, are shown in Fig. 3. According to
the fragmentation model, the mesons are produced symmetrically in both the
hemispheres, so that the distributions over $z$ are described by the expression
$D(z)\cdot \sigma_{b \bar b}/2$ at $p_z > 0$ as well as at
$p_z < 0$.  However, one can see that the real picture of production is more
complex. First, the large enhancement compared to the predictions of the
fragmentation model is found in the region of low $z$ for both the hemispheres.
Second, a significant asymmetry takes place for $z>0.5$. 
Both effects are caused by the following reasons. As was shown in
\cite{yafp}, the first reason is the contribution of diagrams like 
the diagram 20 in Fig. 1. In the leading order of perturbation theory the 
latter contribution corresponds to the "resolved" photon split into
the pair of charmed quarks, so that one of those recombines with the 
produced $\bar b$-quark to form the $B_c$-meson. 
Taking into the ratioof the quark charges
$(Q_c/Q_b)^2=4$, this mechanism causes the increase of $B_c$-production
cross-section in the photon hemisphere. The second reason is the destructive 
interference between the $\bar b$-quark fragmentation and the contribution 
caused by the three-gluon vertices appearing in the splitting of initial gluon 
inside the cone of the latter (see diagrams 13, 14, 17, 18, 22 and 3, 4). 
This is why the essential decrease of $B_c$-production
cross-section occurs in the gluon hemisphere.

The distributions $d\sigma/dp_T$ for the forward and backward
hemispheres ($p_z>0$ and $p_z<0$, respectively)
are presented in Fig. 4 for the $B_c$ state. One can see that these
distributions differ from the ones obtained in the fragmentation model
at low transverse momenta. In the region of high
$p_T$, where the fragmentation mechanism dominates, the deviation 
becomes small, as one expects.

Comparing Figs. 3 and 4, one can draw the conclusion that the difference 
between the production cross-sections in the two hemispheres is most 
significant at large $z$ and low $p_T$.
Our calculations show that the same conclusion is valid for the
production of the excited vector state $B_c^*$.

The considered interactions between the monochromatic beams of
gluons and photons do not occur in a real experiment.
To clarify whether one can observe the asymmetry in 
$ep$ or $\gamma p$ experiments or not,  the process of
$ep$-interaction at HERA ($\sqrt{s}=314$ GeV) is
considered. For this purpose, the cross-section of the gluon-photon
subprocess has been convoluted with the structure functions of both 
the photon in electron and the gluon in proton, respectively.
The equivalent photon parameterization in \cite{pdg} has been chosen as the
photonic structure function.  The CTEQ4 functions have been used for the
gluon distributions \cite{cteq}.

The rapidity distribution normalized to unity is shown in Fig. 5 for the
production of $B_c^*$. For comparison, we have also presented the distribution 
with a constant matrix element, $|A|^2=const$.  One can see that the
differential cross-section for $B_c^*$ production is strongly
displaced in comparison with the distribution calculated with the constant
matrix element. Therefore, the effect of destructive interference, which
has been found at high energies 
of the $g  \gamma  \to  B_c(B_c^*)  +X$ subprocess,
survives in the real conditions of interaction at HERA, too. 
On the other hand, in the fragmentation regime
at high $p_T$ the asymmetry must tend to zero. The latter practically
occurs in Fig. 6, wherein the cross-section
$d\sigma/dy$ at $p_T>30$ GeV is shown in comparison with the 
distribution for the constant matrix element at the same cut off, $p_T>30$ GeV.

Summarizing the results above, one can conclude that the effect 
of the forward-backward asymmetry 
in the $B_c^{(*)}$-meson distribution $d\sigma/dy$ is 
basically related with the 
nonabelian texture of QCD, and this asymmetry is determined by the
three-gluon vertex.

As has been mentioned above, the total cross-section of 
$B_c$ and $B_c^*$ production at the HERA facilities is about to $2.5$ pb 
including charge-conjugate states.
The vector $B_c^*$ state decays to the ground
$B_c$ meson through the emission of a $\gamma$-quantum in the M1-radiative
transition, which has the branching ratio close to 100\% \cite{prspec}.
So,  at the HERA luminosity
${\cal L}=16 \cdot 10^{30}s^{-1}cm^{-2}$ the calculated value of the
cross-section will give only 400 events with the $B_c$ meson per year.  
The increase of total energy
by three times, which correspond to the combination of LEP and LHC
accelerators, will result in the enhancement of the asymmetry as well as the
total cross-section, which is estimated to have a value of 6 nb at
$\sqrt{s}=660$ GeV (see Fig. 7). At the luminosity $10^{32}s^{-1}cm^{-2}$
one can expect about $6\cdot 10^3$ $B_c^{(*)}$ events per year.

Taking into account low branching fractions for the exclusive
decay modes of $B_c$ (for example, $Br(B_c^+ \to J/\psi e^+ \nu ) 
\simeq 2.7 \%$)  and the efficiency for the detecting of decay process
($\epsilon \simeq 0.1$), one can see that the expected number of events
can turn out to be insufficient for the investigation 
of the described asymmetry in the $ep$-production of $B_c$ meson. 

The study of these nonabelian effects
is a more real problem for the production of $D_s$ meson. 
As we can see above, the asymmetry is caused by quite similar deviation 
from the fragmentation mechanism due to the process-dependent contributions,
which have a more rapid decrease with the growth of the transverse momentum
of the produced meson. For the $B_c^{(*)}$ mesons these sub-leading terms
can be reliably calculated in perturbative QCD. In the $D_s$ production,
higher orders in $\alpha_s$ can be significant. Nevertheless, we believe that
the sub-leading non-fragmentational contributions have the same feature as
in the $B_c$ production: the yield in the photon hemisphere is enhanced
and the one in the gluon hemisphere is suppressed due to the nonabelian
interactions of gluons.
One has to note that
the quark-mass ratio determining the asymmetry value in the model 
under consideration is one and the same for the $D_s$ and $B_c$ mesons,
and hence the effect of destructive interference can be experimentally
observed at HERA, where the energy scale is given by the ratio of
$M_{D_s}/\sqrt{s}$, which is approximately equal to
$M_{B_c}/\sqrt{s}$ at the energy of LEP and LHC combination.
The distribution for the $D_s$-meson rapidity at the HERA energy 
is shown in Fig. 8.

A more detailed study of the interference effect
requires the investigation of the contributions by individual diagrams in a
proper gauge. The latter consideration will be presented elsewhere.

\vspace*{3mm}
This work is in part supported by the Russian Foundation for Basic Research,
grants 96-02-18216 and 96-15-96575. V.V.K. and A.K.L. express 
their gratitude to Profs.A.Wagner and P.M.Zerwas for  
kind hospitality and support during the visit to DESY, where this work was
completed. 

\pagestyle{empty}

\parbox{16cm}{%
\vspace*{2cm}
\hspace*{0.5cm}\epsfxsize=14cm \epsfbox{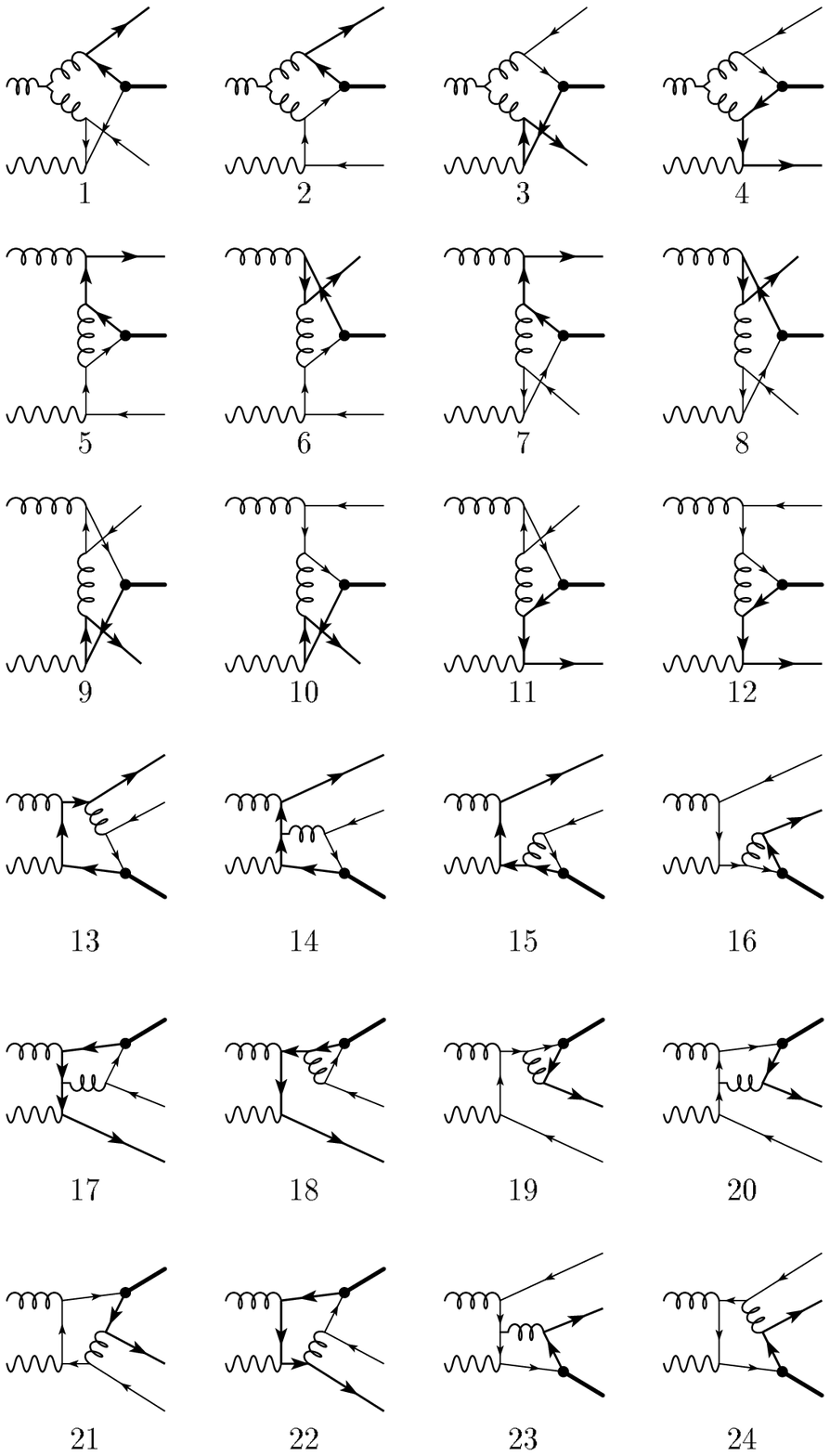}

\vspace*{-3.cm}
\mbox{\bf Fig. 1.}
The diagrams contributing to the gluon-photon production of
$(\bar b c)$-quarkonium in the leading order of perturbative theory.
The $c$-, $b$-quarks, gluons and photons are denoted by the thin, thick,
spiral and wavy lines, respectively.
}
 
\newpage
\parbox{16cm}{%
\vspace*{1.cm}
\hbox to 1.5cm {\hfil\mbox{$\sigma_{g\gamma}^{B_c^{(*)}}$, pb}}
\vspace*{0.5cm}
\epsfxsize=14cm \epsfbox{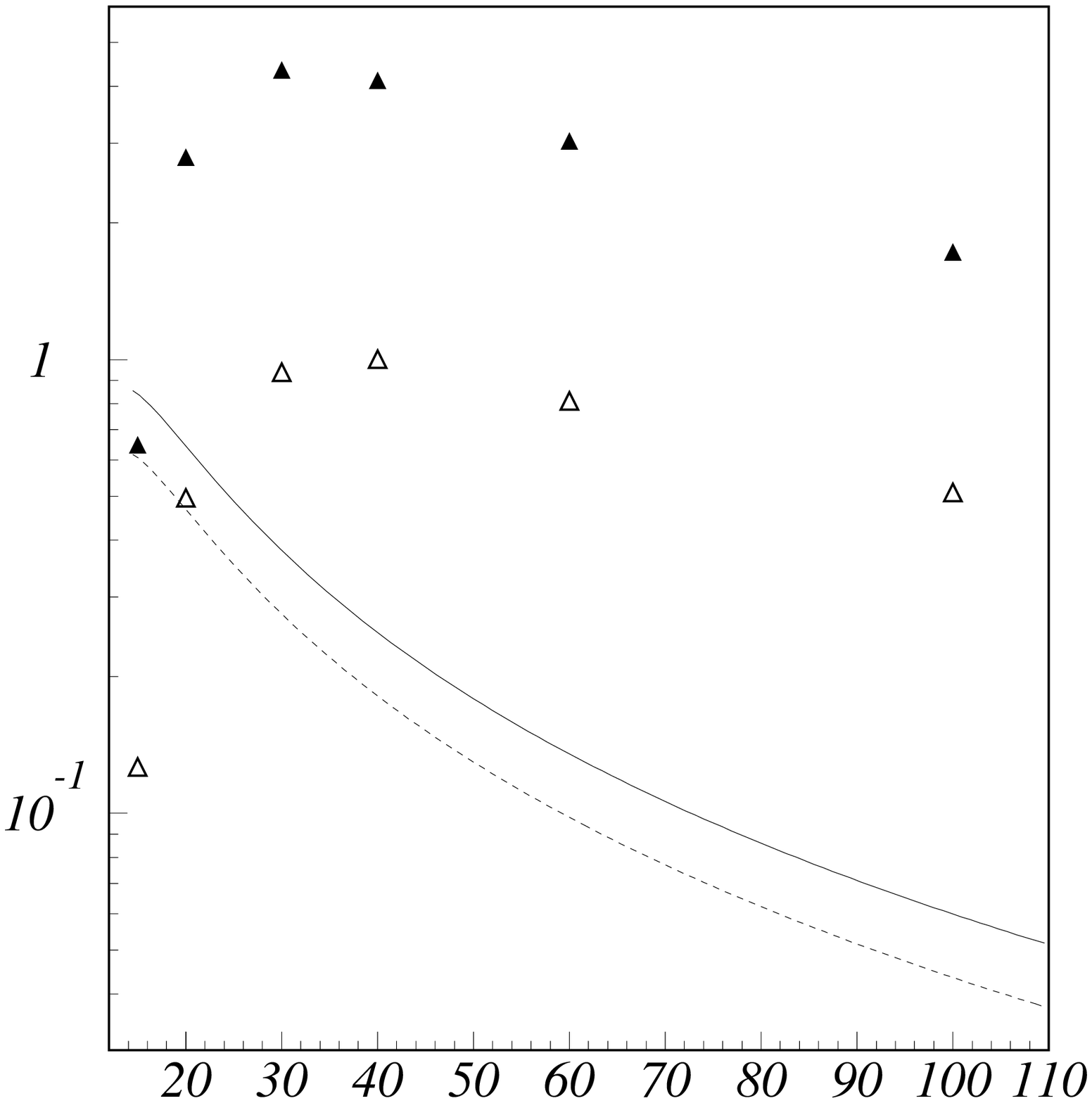}

\vspace*{-0.5cm}
\hbox to 16.cm {\hfil \mbox{$\sqrt{s_{g\gamma}}$, GeV}}
\vspace*{1.cm}
\mbox{\bf Fig. 2.}
The dependence of the gluon-photon cross-section for the production of
$B_c^*$ (solid triangle) and $B_c$ (empty triangle) in comparison with
the predictions of the fragmentation mechanism (solid and dashed curves,
respectively).
}

\newpage
\parbox{16cm}{%
\vspace*{1.cm}
\hbox to 1.5cm {\hfil\mbox{
$d\sigma^{B_c}_{g\gamma}/d z$, pb}}
\vspace*{0.5cm}
\epsfxsize=14cm \epsfbox{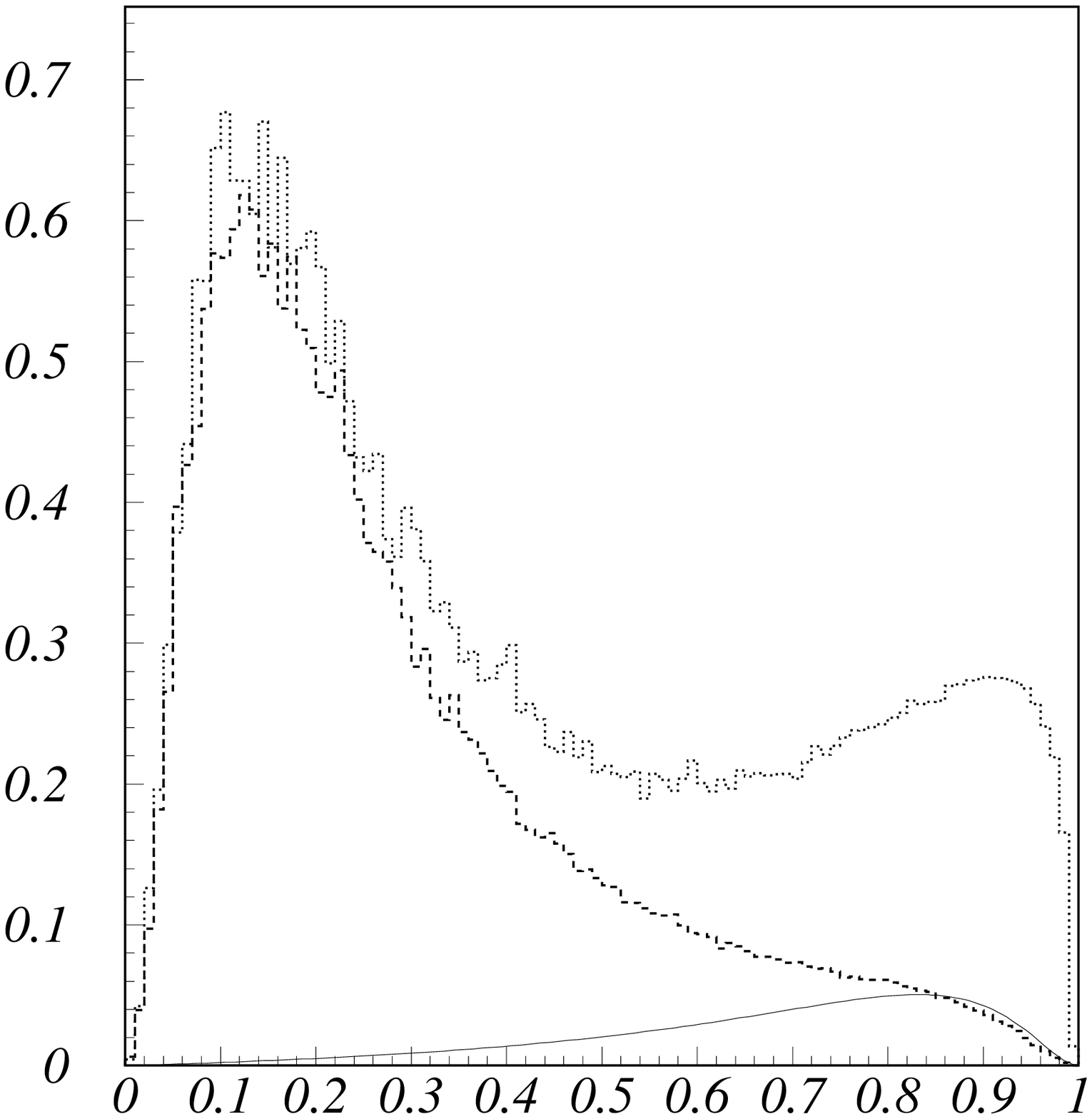}

\vspace*{-0.9cm}
\hbox to 16.cm {\hfil \mbox{$Z$\hspace*{5cm}}}
\vspace*{1.cm}
\mbox{\bf Fig. 3.}
The $d\sigma^{B_c}_{g\gamma}/d z$ for the $B_c$ meson moving in the gluon
hemisphere (dashed histogram) and for the one moving in the photon hemisphere
(dotted histogram) at the energy of gluon-photon interaction equal to
100 GeV. For the sake of comparison, the prediction of the 
fragmentation model is also presented (solid curve).
}

\newpage
\parbox{16cm}{%
\vspace*{1.cm}
\hbox to 1.5cm {\hfil\mbox{
$d\sigma^{B_c}_{g\gamma}/d p_T$, pb/GeV}}
\vspace*{0.5cm}
\epsfxsize=14cm \epsfbox{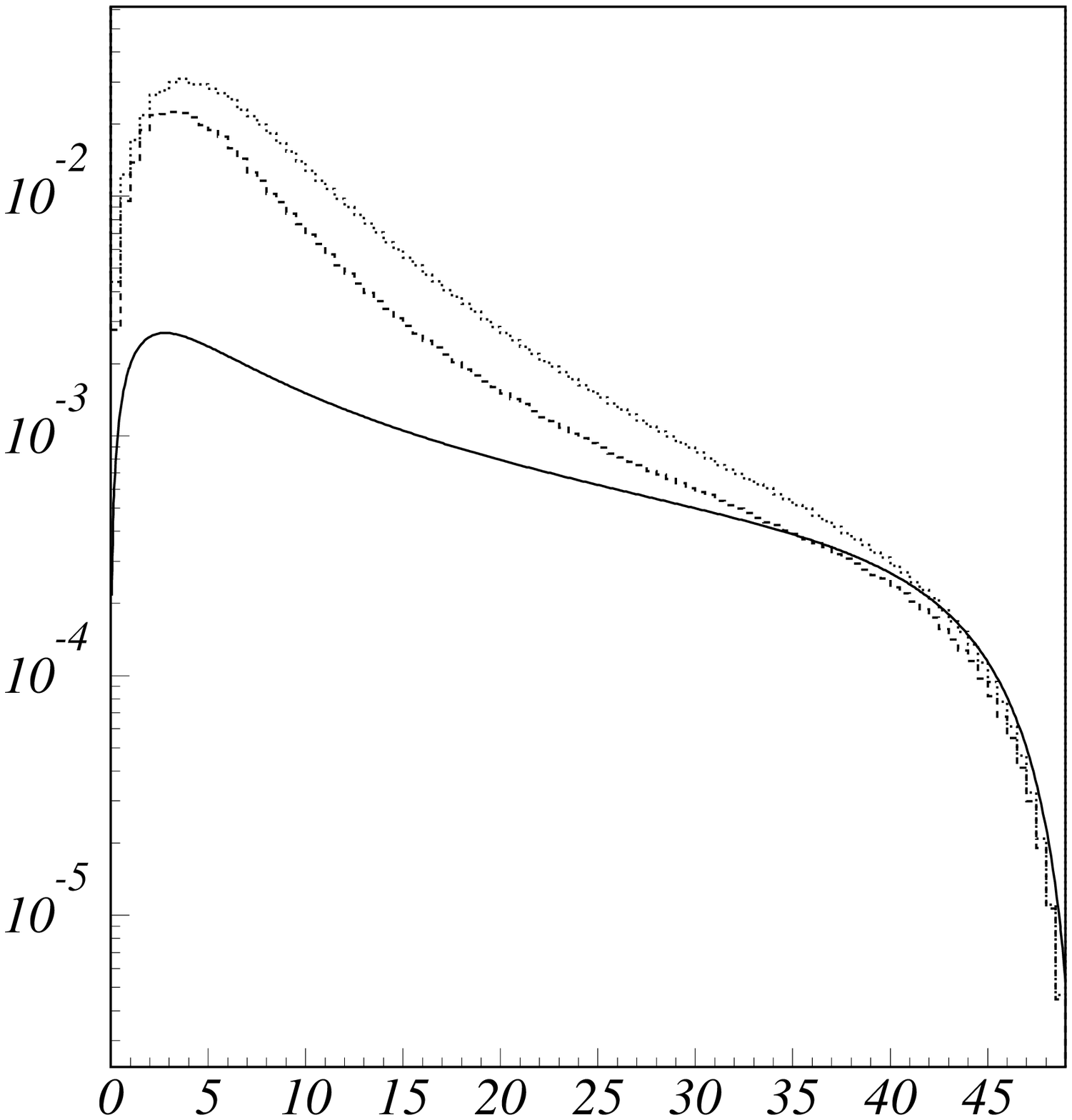}

\vspace*{-0.5cm}
\hbox to 16.cm {\hfil \mbox{$p_T$, GeV\hspace*{2cm}}}
\vspace*{1.cm}
\mbox{\bf Fig. 4.}
The differential cross-section for the gluon-photon production of
$B_c$ versus the transverse momentum at the energy of 100 GeV.
The notations are the same as in Fig. 3.
}

\newpage
\parbox{16cm}{%
\vspace*{1.cm}
\hbox to 1.5cm {\hfil\mbox{
$\sigma^{-1}d\sigma^{B_c^*}_{ep}/d y$}}
\vspace*{0.5cm}
\epsfxsize=14cm \epsfbox{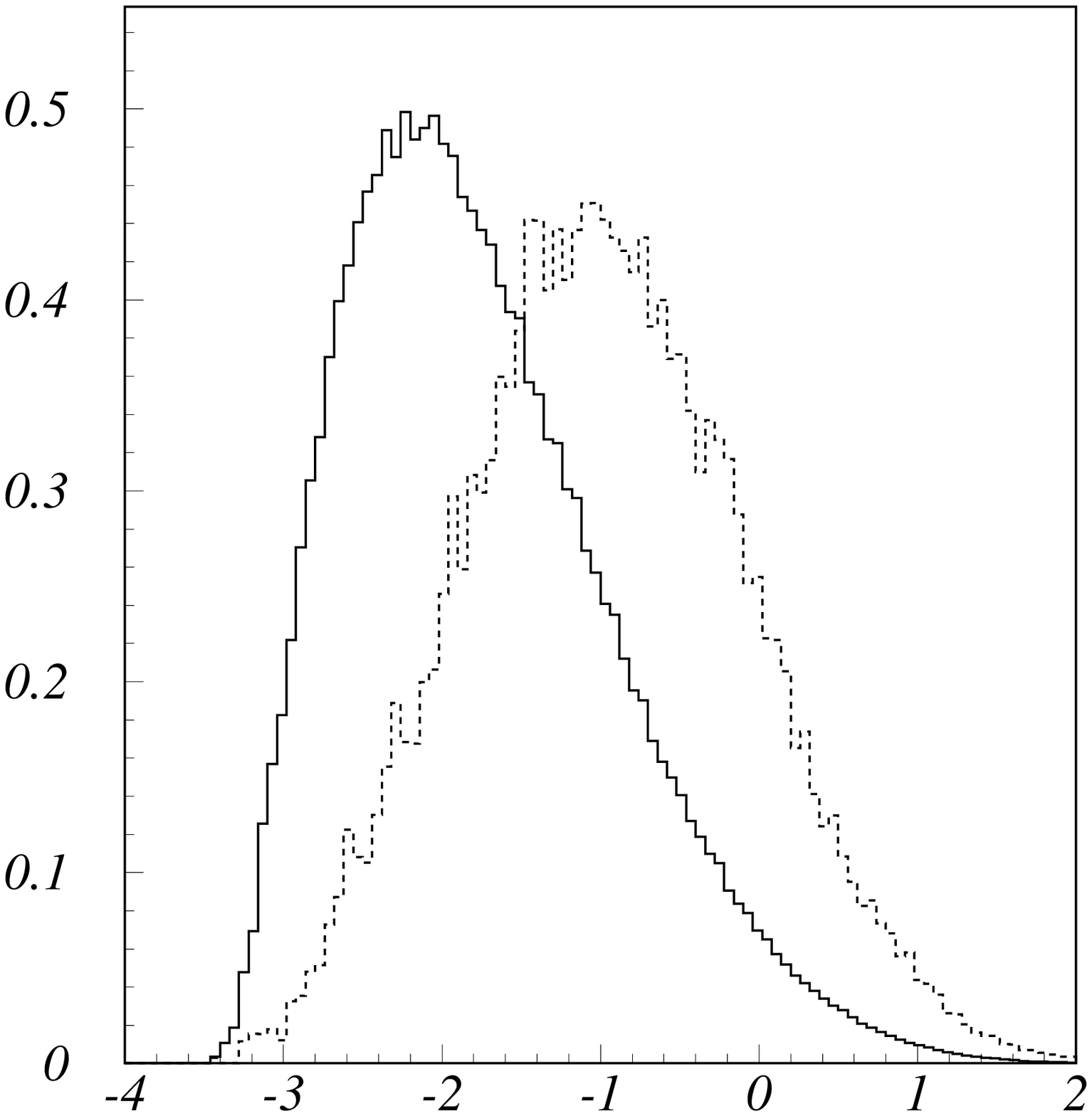}

\vspace*{-0.5cm}
\hbox to 16.cm {\hfil \mbox{$y$\hspace*{2cm}}}
\vspace*{1.cm}
\mbox{\bf Fig. 5.}
The differential cross-section normalized to unit area over the rapidity of
$B_c^*$ produced in $ep$-interactions at $\sqrt{s}=314$ GeV (solid histogram).
For the sake of comparison, the normalized distribution with the constant
matrix element is shown by the dashed histogram. The rapidity $y$ is 
defined in the c.m.s. of the colliding beams.
}

\newpage
\parbox{16cm}{%
\vspace*{1.cm}
\hbox to 1.5cm {\hfil\mbox{
$\sigma^{-1}d\sigma^{B_c^*}_{ep}/d y (p_T>30 {\rm \ GeV}$) }}
\vspace*{0.5cm}
\epsfxsize=14cm \epsfbox{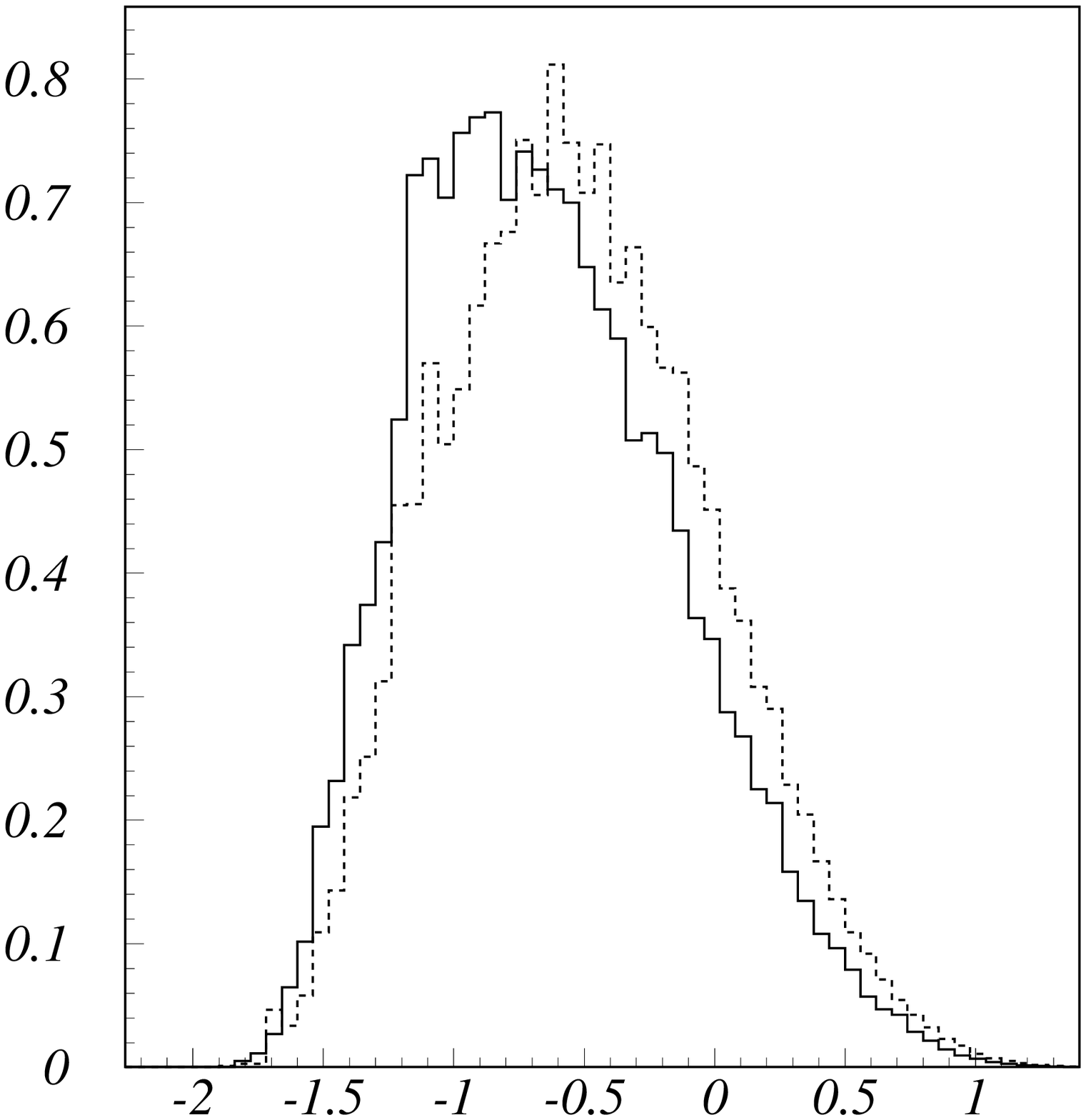}

\vspace*{-0.5cm}
\hbox to 16.cm {\hfil \mbox{$y$\hspace*{2cm}}}
\vspace*{1.cm}
\mbox{\bf Fig. 6.}
The same as in Fig. 5, but with $p_T> 30$ GeV. 
}

\newpage
\parbox{16cm}{%
\vspace*{1.cm}
\hbox to 1.5cm {\hfil\mbox{
$\sigma^{-1}d\sigma^{B_c^*}_{ep}/d y$}}
\vspace*{0.5cm}
\epsfxsize=14cm \epsfbox{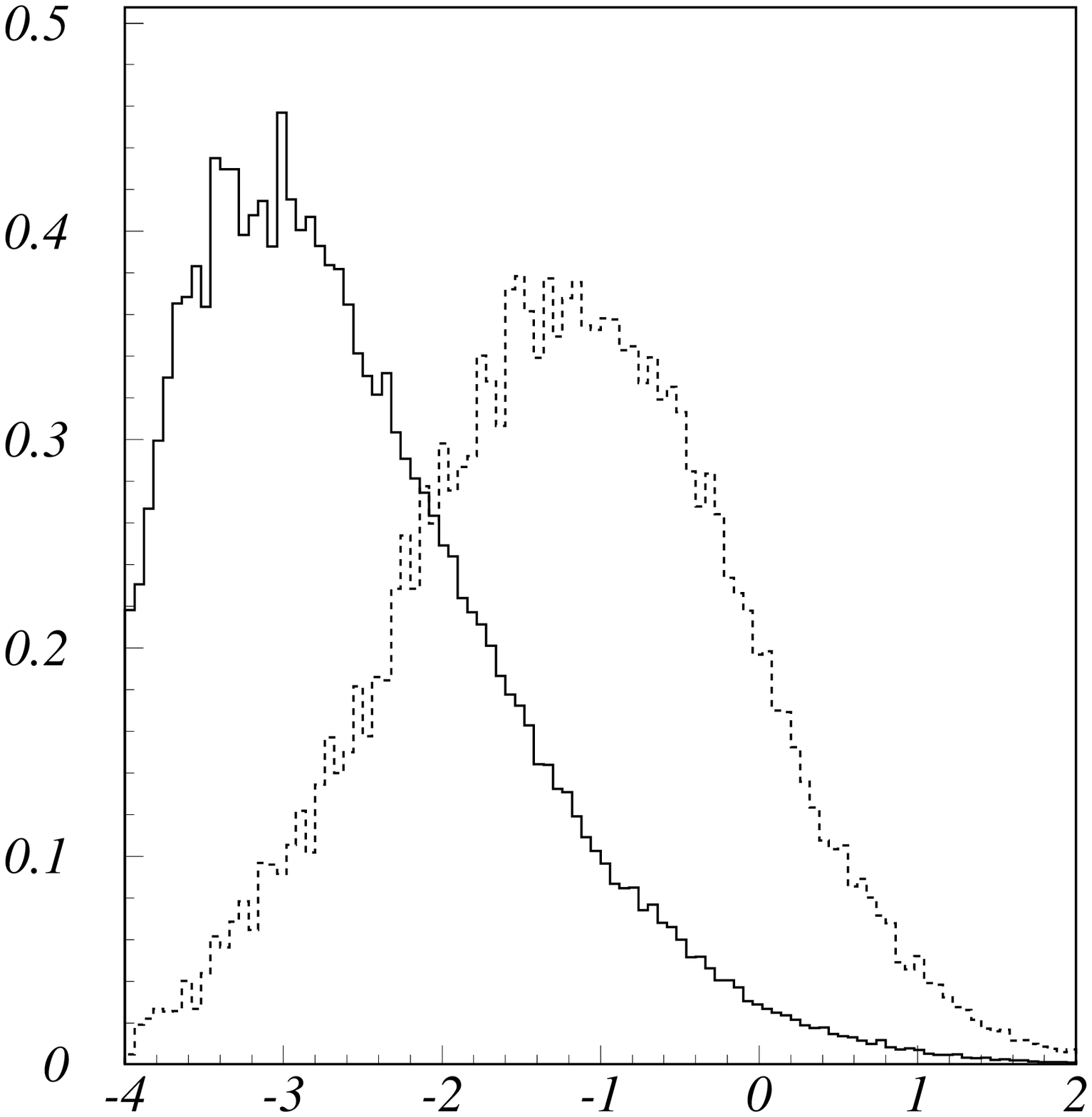}

\vspace*{-0.5cm}
\hbox to 16.cm {\hfil \mbox{$y$\hspace*{2cm}}}
\vspace*{1.cm}
\mbox{\bf Fig. 7.}
The same as in Fig. 5, but with the energy of $ep$-interactions,
$E=660$ GeV.
}

\newpage
\parbox{16cm}{%
\vspace*{1.cm}
\hbox to 1.5cm {\hfil\mbox{
$\sigma^{-1}d\sigma^{D_s^*}_{ep}/d y$}}
\vspace*{0.5cm}
\epsfxsize=14cm \epsfbox{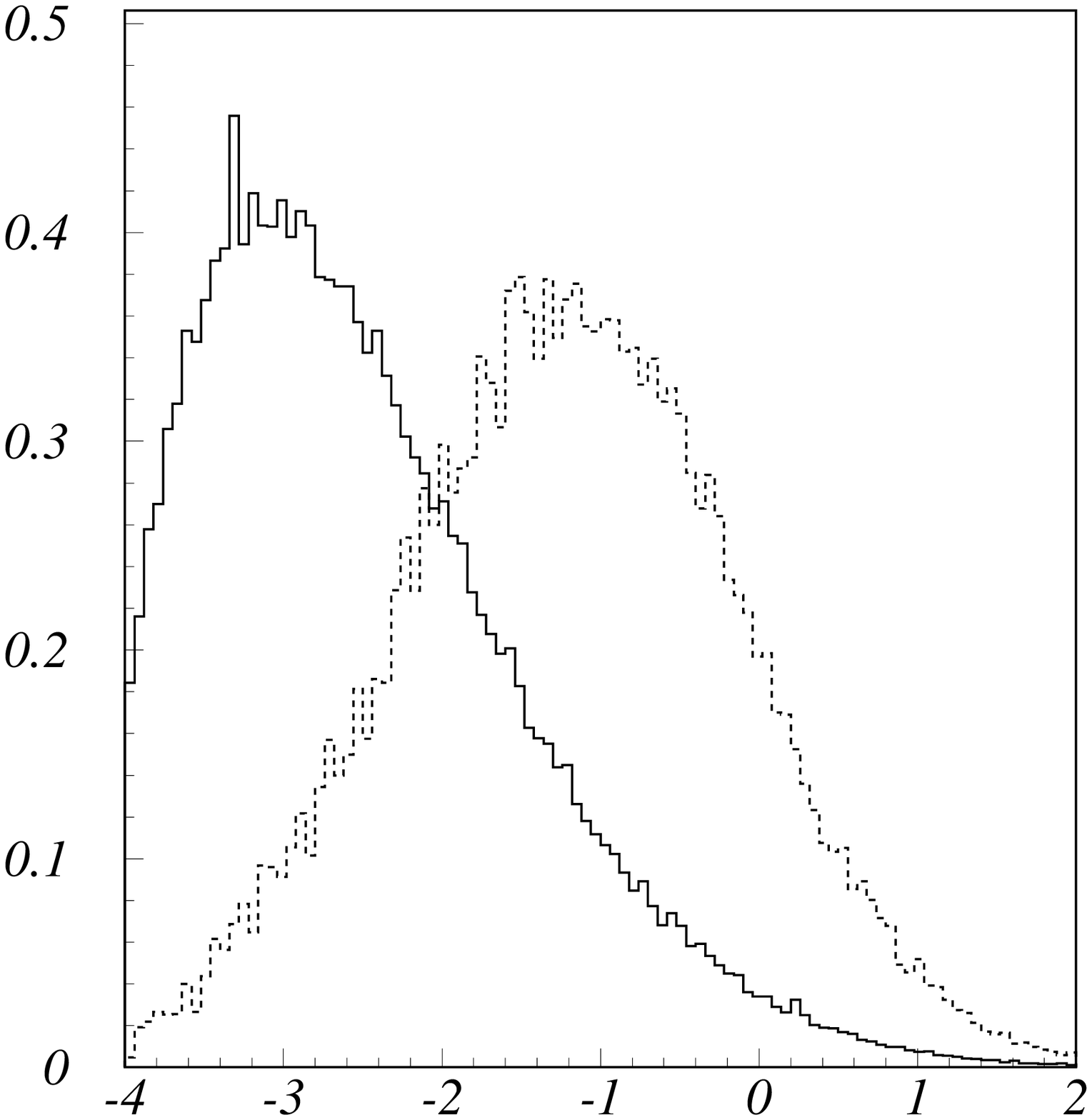}

\vspace*{-0.5cm}
\hbox to 16.cm {\hfil \mbox{$y$\hspace*{2cm}}}
\vspace*{1.cm}
\mbox{\bf Fig. 8.}
The differential cross-section normalized to unit over the rapidity of
$D_s^*$ produced in $ep$-interactions at $\sqrt{s}=320$ GeV (solid histogram).
For the sake of comparison, the normalized distribution with the constant
matrix element is shown by the dashed histogram.
}

\end{document}